\documentclass{osa-article}
\usepackage[utf8]{inputenc} 
\usepackage{graphicx}
\usepackage{float}
\usepackage{siunitx}
\usepackage{fdsymbol}
\usepackage{array}
\usepackage{subcaption}
\journal{oe}

\begin{document}

\title{Erbium-doped aluminophosphosilicate all-fiber laser operating at 1584 nm}

\author{Marie-Pier Lord,\authormark{*} Lauris Talbot, Olivier Boily, Tommy Boilard, Guillaume Gariépy, Sacha Grelet, Pascal Paradis, Nicolas Grégoire, Steeve Morency, Younès Messaddeq and Martin Bernier}

\address{Centre d'optique, photonique et laser (COPL), Université Laval, Québec, G1V 0A6, Canada}

\email{\authormark{*}marie-pier.lord.1@ulaval.ca}

\begin{abstract}
We report on an ytterbium-free erbium-doped aluminophosphosilicate all-fiber laser, producing an output power of 25 W at a wavelength of 1584 nm with a slope efficiency of 30\% with respect to the 976 nm absorbed pump power. The simple cavity design proposed takes advantage of fiber Bragg gratings written directly in the gain fiber. The single-mode erbium-doped aluminophosphosilicate fiber was fabricated in-house and was doped with 0.06 mol.\% of Er$_{2}$O$_{3}$, 1.77 mol.\% of Al$_{2}$O$_{3}$ and  1.04 mol.\% of P$_{2}$O$_{5}$. The incorporation of aluminium and phosphorus into the fiber core allowed for an increased concentration of erbium without inducing significant clustering, while keeping the numerical aperture low to ensure a single-mode laser operation. 
\end{abstract}

\section{Introduction}
The $^4\text{I}_{13/2} \rightarrow ^4\text{I}_{15/2}$ transition in Er$^{3+}$ ions can be exploited for a wide variety of applications, as its corresponding emission around 1.6 µm matches the "eye-safe" region of the electromagnetic spectrum \cite{ICNIRP:13} as well as the high-transmission window of both silica fibers and the atmosphere. Among such applications belong light detection and ranging (LIDAR) \cite{Frehlich:97}, spectroscopy and optical wireless communication systems \cite{Elgala:11}. Laser sources emitting around 1.6 µm can also be used to pump cavities that emit further in the infrared region of the spectrum \cite{Solodyankin:08}. Monolithic all-fiber lasers are greatly suited to these functions as they provide high quality beams through robust and compact setups. It has however been proven difficult for such lasers to achieve high efficiency for direct diode pumping schemes, mainly due to the corresponding small absorption cross section of erbium as well as the occurrence of concentration quenching at high concentration \cite{Delevaque:93}. To date, the most powerful diode-pumped erbium-doped silica fiber laser was demonstrated by Lin et al., who reported on a 656 W erbium-doped fiber laser yielding an efficiency of 35.6\% with respect to the launched pump power at 0.98 µm \cite{Lin:18}. However, the setup used was not all-fiber and used a multimode fiber yielding a poor beam quality (M$^2\approx10.5$).

One of the commonly used solutions to increase pump absorption in erbium-doped fiber lasers is to integrate ytterbium as a codopant in the fiber core. For instance, Jebali et al. demonstrated an in-band pumped monolithic Er-Yb codoped fiber laser reaching 264 W of output power at 1584 nm with an efficiency of 74\% with respect to the 1525 nm pump power \cite{Jebali:14}. Similarly, Jeong et al. reported on a 297 W Er-Yb codoped fiber laser emitting at 1567 nm and yielding an efficiency of 21\% with respect to the launched pump power at 975 nm \cite{Jeong:07}. The incorporation of ytterbium as a codopant has, however, the effect of increasing the refractive index of the core, meaning the beam quality can be compromised if the core area is not reduced. Besides, parasitic emission around 1 µm coming from Yb$^{3+}$ ions frustrates the use of such lasers \cite{Jeong:07}.

Another method to increase laser performances, by reducing clustering of erbium ions in the fiber core, is to use aluminum or phosphorous as a codopant \cite{Scrivener:89,Vienne:98}. Those elements, either by forming a solvation shell around rare earth ions - as is the case for phosphorous - or by increasing the entropy change of mixing - as is the case for aluminum - hinder the formation of clusters and prevent the emergence of detrimental pair-induced quenching \cite{Funabiki:12}. However, several problems may arise from the use of such codopants, like an undesired increase of the refractive index of the core, or radiation induced attenuation \cite{Savelii2017,Likhachev:09}.

A method to overcome those limitations is to use both aluminum and phosphorous as codopants in equimolar proportions. By doing so, AlPO\textsubscript{4} structures are specifically formed inside the fiber core \cite{aboud2015}. Those units have a structure similar to that of pure silica and thus preserve the refractive index of the core closest to that of the pure silica cladding, which allow for obtaining low-NA doped fibers. Furthermore, the formation of AlPO\textsubscript{4} units allows an increased concentration of rare-earth dopants, owing to the ability of strongly polarized Al-O-P bonds to screen the positive charge of rare-earth ions \cite{Likhachev:09}. With this technique, in conjunction with the high efficient Yb system, Liu et al. achieved a 3.03 kW laser output with a slope efficiency of 76.6\% at 1080 nm when pumping at 976 nm \cite{Liu:18}. This ytterbium-doped laser provides a near single-mode beam quality (M$^2\approx1.58$) and a low numerical aperture (NA $\approx$ 0.065). Jebali et al. have also successfully applied this codoping technique in the case of their high-power erbium-ytterbium laser \cite{Jebali:14}. This approach has also been applied for an ytterbium-free erbium fiber laser by Kotov et al., who demonstrated a record-high laser slope efficiency of 40\%, yielding an output power limited to 7.5 W \cite{kotov2012high}. In this demonstration, the fiber had a squared-shape cladding of 80 µm in diameter, which imposed the use of a tapered fiber to efficiently couple the pump beam, which limits the power scalability. Also, the fiber used was co-doped with 1.5 mol.\% of germanium-oxide, in addition to 9 mol.\% of both Al$_2$O$_3$ and P$_2$O$_5$, and 0.1 mol.\% of Er$_2$O$_3$.

In this paper, we present the laser emission capabilities of an erbium-doped aluminophosphosilicate (Er$^{3+}$- APS) all-fiber laser operating at 1584 nm, with an appropriate design for power scaling. Pumped by a commercial 976 nm laser diode and bounded by a pair of intracore fiber Bragg gratings (FBGs) written directly in the gain fiber, the cavity proposed provides a single-mode output signal, yielding 30\% of efficiency with respect to the absorbed pump power. Numerical modeling confirms that the integration of aluminium and phosphorus into the fiber core significantly reduce the clustering and thus restrain the event of concentration quenching.

\section{Fiber fabrication and properties}

A silica glass preform was fabricated in-house using the MCVD technique. A porous silica layer was deposited on the inner surface of a substrate tube (Suprasil-F300, outer diameter 25 mm, inner diameter 19 mm, from Heraeus). Doping was performed with the solution doping process, using erbium and aluminum inorganic salts as precursors (ErCl\textsubscript{3}$\smblkcircle$6H\textsubscript{2}O, 99.999\%, from MV Laboratories,  and AlCl\textsubscript{3}$\smblkcircle$6H\textsubscript{2}O, 99.9999\%, from Pure Analytical Laboratories), along with phosphoric acid (85\% H\textsubscript{3}PO\textsubscript{4}, 99.999\%, from Sigma Aldrich) and high purity deionized water (18.15 M$\Omega$).

Phosphorous and aluminum were incorporated in appropriate quantities in order to obtain an almost equimolar ratio inside the fiber's core. An equimolar proportion of aluminum and phosphorous favors the formation of AlPO\textsubscript{4} structures. An excess of either aluminum or phosphorous, while not being strictly detrimental to the fiber's performances, is not as efficient to limit the formation of erbium clusters \cite{Likhachev:09}.

The preform was polished on two sides into a "double D" shape to favor pump absorption by inhibiting the propagation of poorly interactive circular modes \cite{Pleau:18}. The cladding to core aspect ratio was optimized to improve pump absorption while retaining a single-mode operation at the wavelength of emission (i.e. 1584 nm).

\begin{figure}[H]
    \centering
    \includegraphics[trim={3cm 2cm 4cm 2cm},clip, scale=0.08]{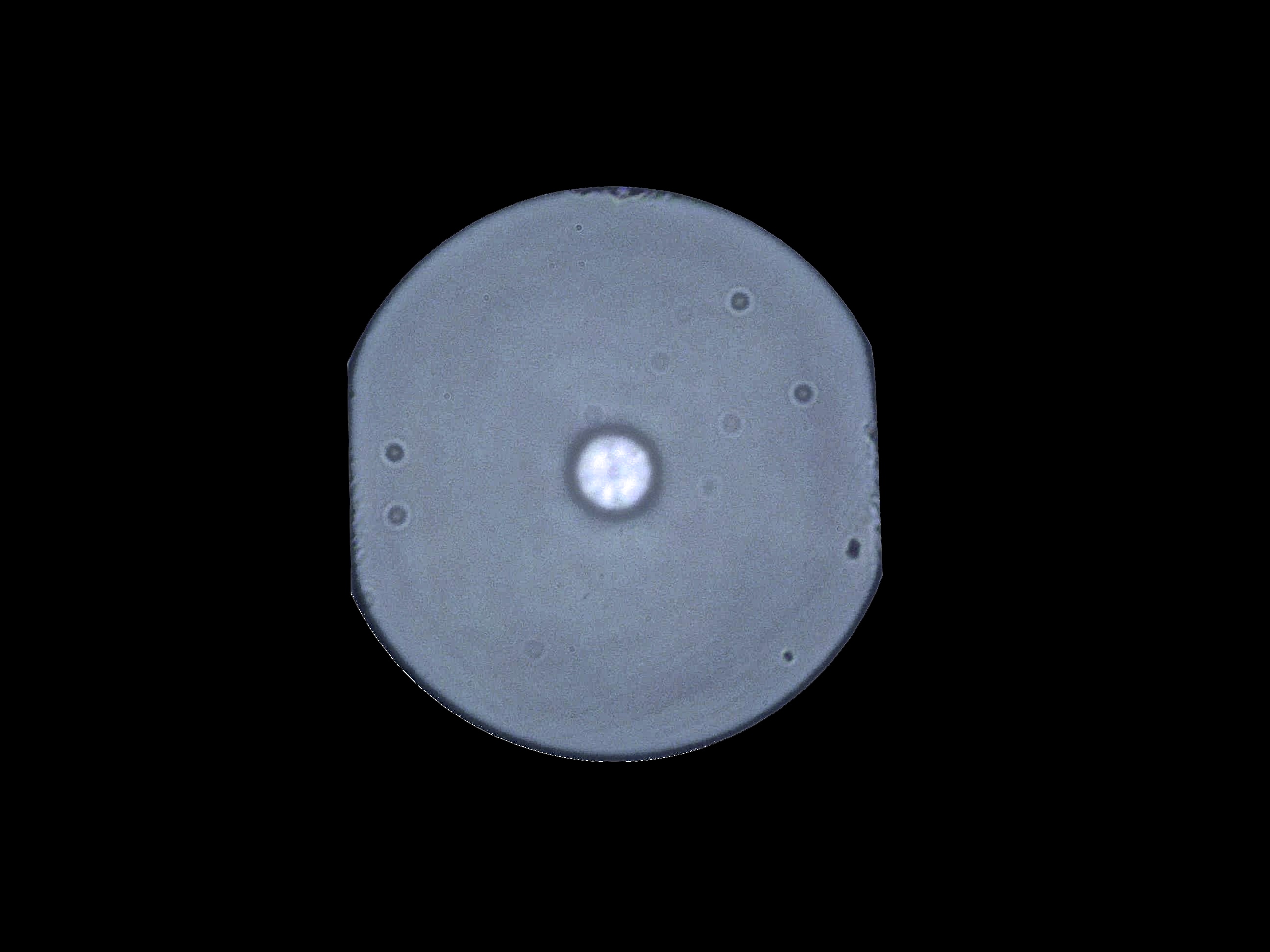}
    \caption{Optical microscope picture of the cross-section of the aluminophosphosilicate fiber.}
    \label{fig:cross}
\end{figure}

Fiber characteristics are summarized in table 1 and are compared to our previous work, while the fiber cross-section is depicted in Fig. \ref{fig:cross}. Cutoff values were measured using the bend reference technique, attenuation measurements were obtained via cutback, and the numerical aperture was calculated from the measured cutoff and core size values. The percentage of paired ions was estimated by the numerical model described in section \ref{s:3} hereafter.

\begin{table}[H]
\captionsetup{font=bf}
\caption{Parameters of the APS fiber compared to our previous work.}
\begin{tabular}{lrr}
  \hline
  Parameter  & Er$^{3+}$- Al (Pleau et al., \cite{Pleau:18})  & Er$^{3+}$- APS (this work)\\
  \hline
  Core Diameter [µm]   & 16 & 17\\
  Cladding diameter [µm] & 120 x 130  & 120 x 130\\
  Cutoff wavelength [nm]   & 1480 & 1369\\
  Numerical aperture & 0.07  &0.06 \\
  Er\textsubscript{2}O\textsubscript{3} concentration [mol.\%] & 0.03  &0.06\\
  Al\textsubscript{2}O\textsubscript{3} concentration [mol.\%] & 1.2 & 1.77\\
  P\textsubscript{2}O\textsubscript{5} concentration [mol.\%] & -  & 1.04\\
  Clad absorption at 976 nm [dB/km]  & 43.5 & 199\\
  Core loss at 1.1 µm [dB/km] & 23.7 & 8.9 \\
  Estimated \% of paired ions (2\emph{k}) & 9.3  & 4.2\\
  Optimal length [m]  & 60 & 28.5\\
  \hline
\end{tabular}
\label{tab:1}
\end{table}

It can be observed in table \ref{tab:1} that clustering is significantly reduced in the aluminophosphosilicate fiber - the amount of paired ions remaining around 4\% - even for a relatively high erbium ion concentration of 0.06 mol.\%. In comparison with our previous work \cite{Pleau:18}, erbium concentration is doubled while clustering is halved and the numerical aperture is kept low, which is enabled by the aluminum-phosphorous codoping.

\begin{figure}[H]
\begin{subfigure}{.5\textwidth}
	\centering
\includegraphics[scale=0.45]{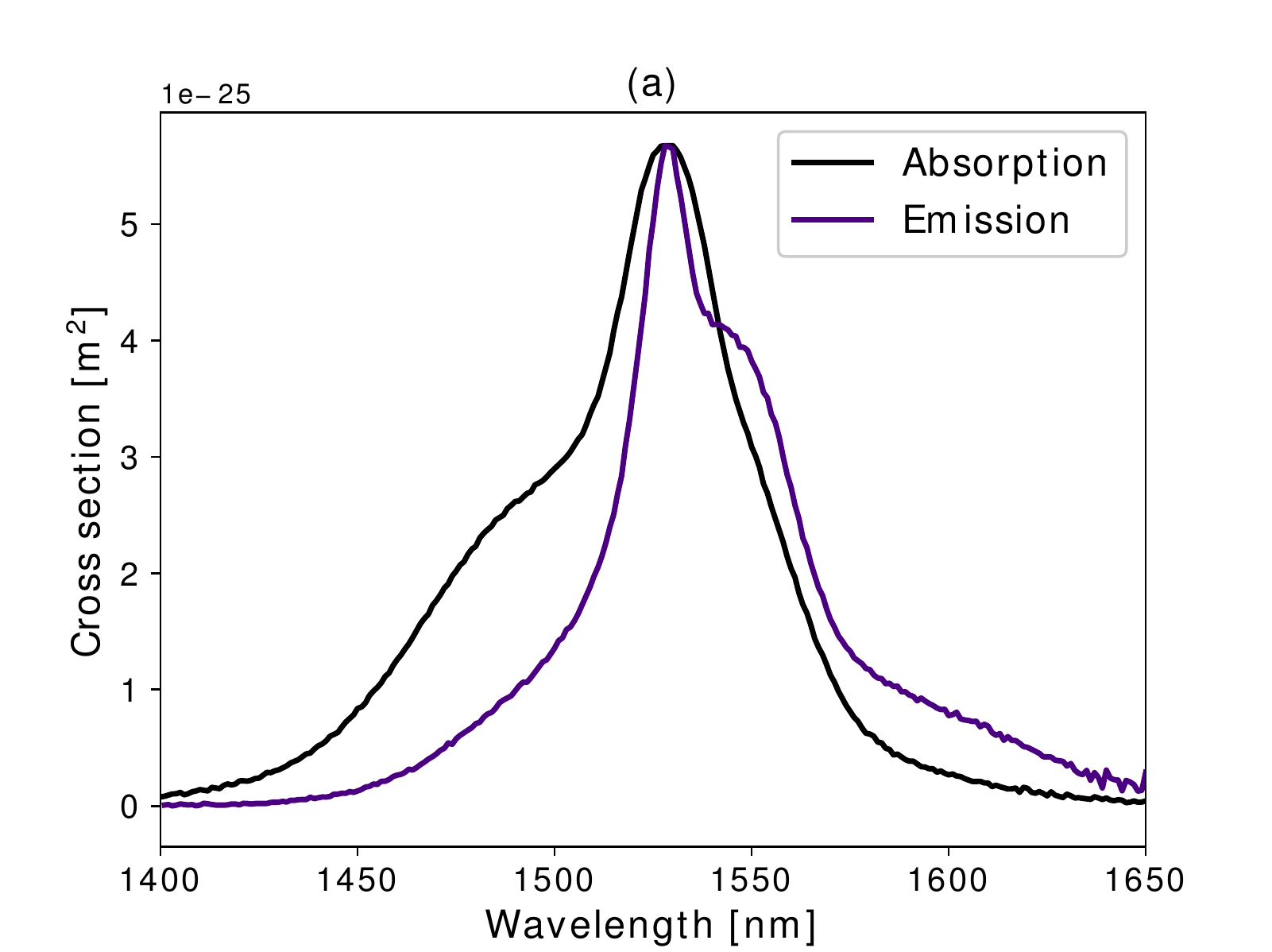}
\end{subfigure}
\begin{subfigure}{.5\textwidth}
	\centering
\includegraphics[scale=0.45]{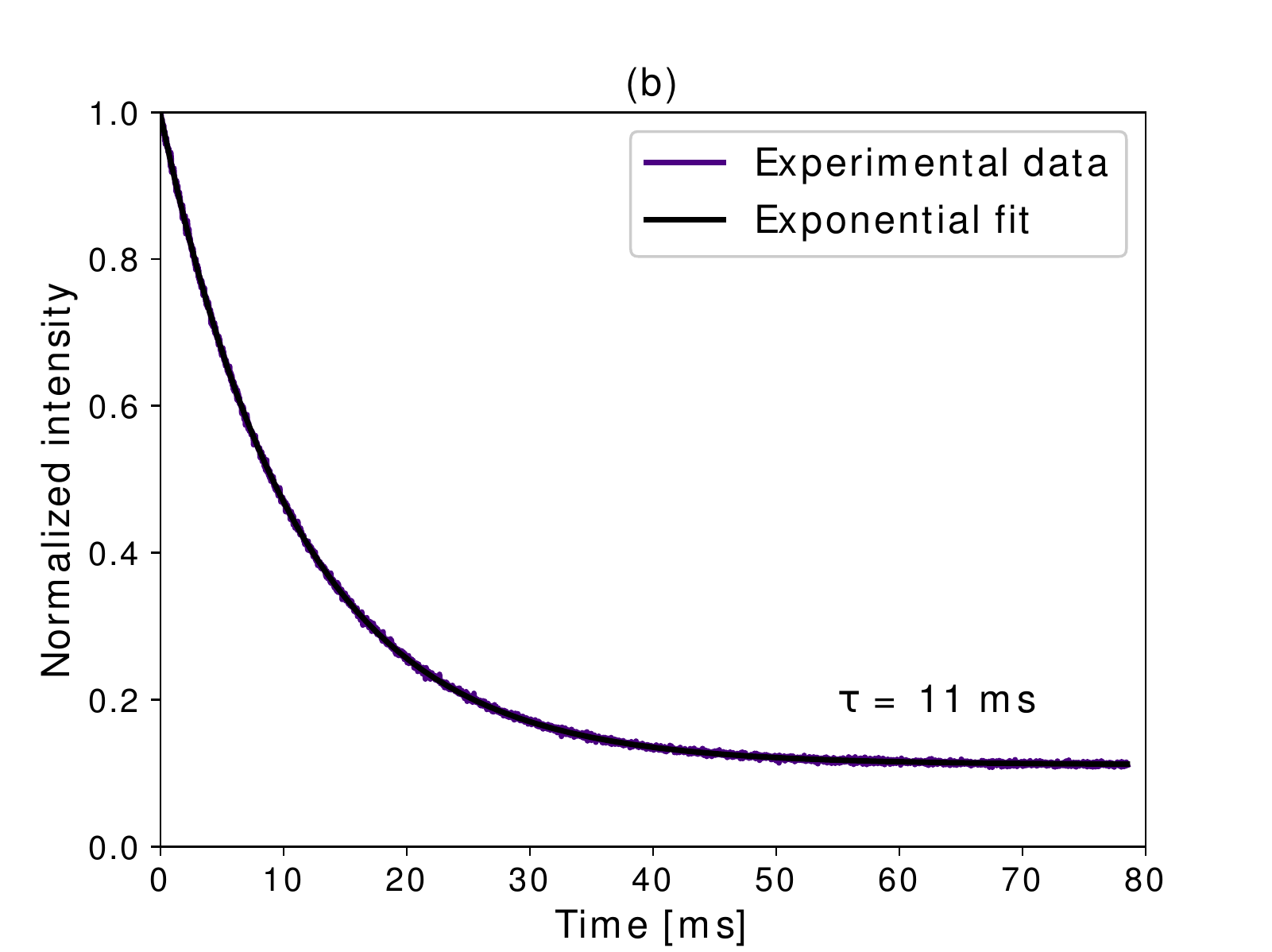}
\end{subfigure}
\caption{(a) Measured cross section of absorption and emission around the signal wavelength. (b) Measured luminescence decay of the $^4\text{I}_{13/2}$ excited level.}
\label{fig:fibre}
\end{figure}

The emission and absorption cross sections of the $^4\text{I}_{13/2} \rightarrow ^4\text{I}_{15/2}$ transition of the Er$^{3+}$ ions in the APS fiber were measured using the fiber preform and are presented alongside the lifetime measurement of the $^4\text{I}_{13/2}$ level in Fig. \ref{fig:fibre}. The emission cross section and the lifetime measurements were obtained using a Jobin-Yvon Nanolog combined with a Hamamatsu photon counter, while the absorption cross section curve was obtained using a Cary 5000 UV-Vis-NIR made by Agilent.

The results of Fig. \ref{fig:fibre} suggest that the incorporation of aluminium and phosphorus in the fiber core do not substantially change the lifetime of the $^4\text{I}_{13/2}$ excited level or the cross sections of Er$^{3+}$ ions, since the curves follow what has been previously documented in literature for Er$^{3+}$ ions in pure silica glass \cite{Barnes:91}. 

\section{Numerical modeling}
\label{s:3}
\begin{figure}[H]
\begin{subfigure}{.5\textwidth}
	\centering
\includegraphics[scale=0.45]{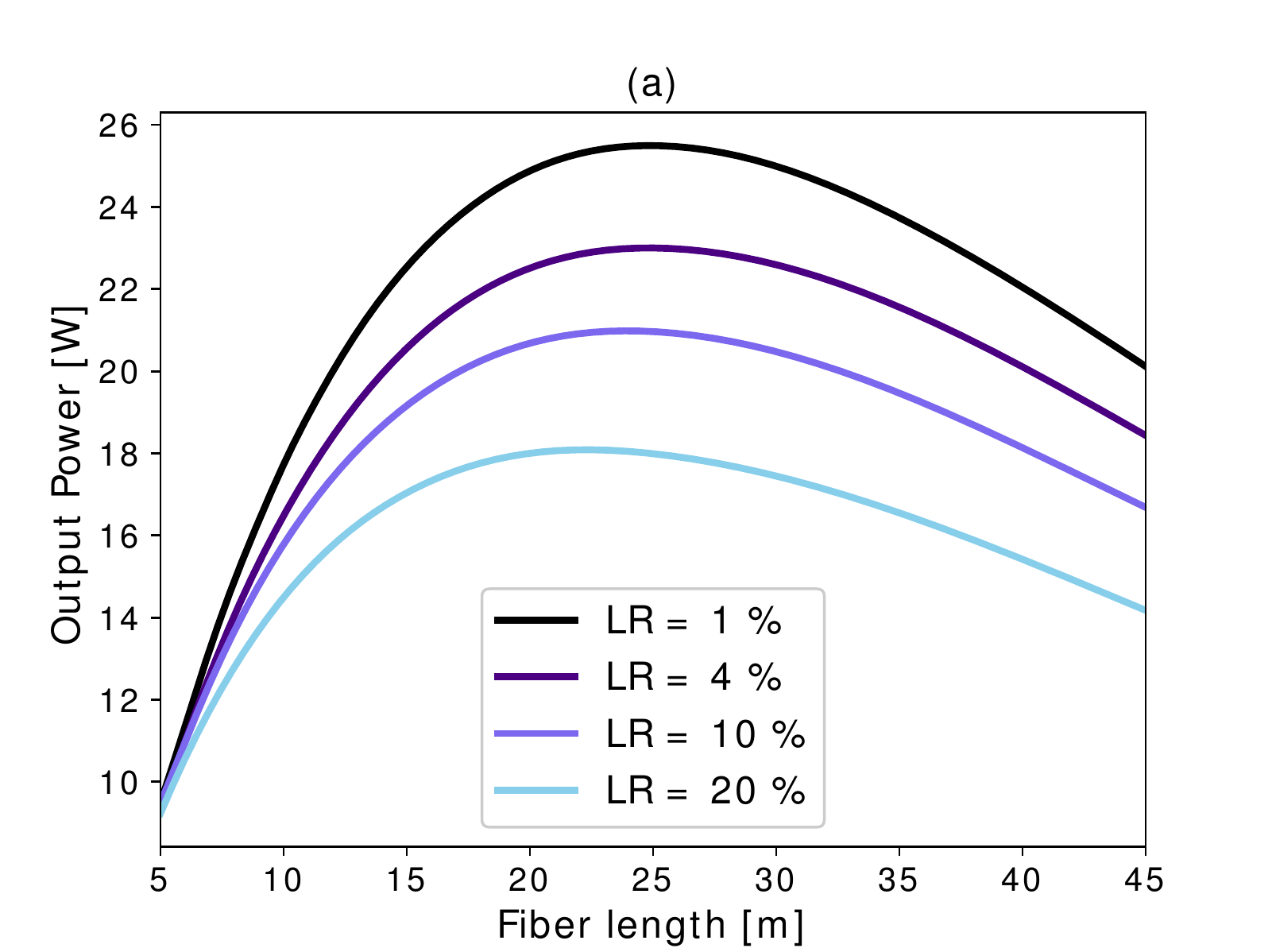}
\end{subfigure}
\begin{subfigure}{.5\textwidth}
	\centering
\includegraphics[scale=0.45]{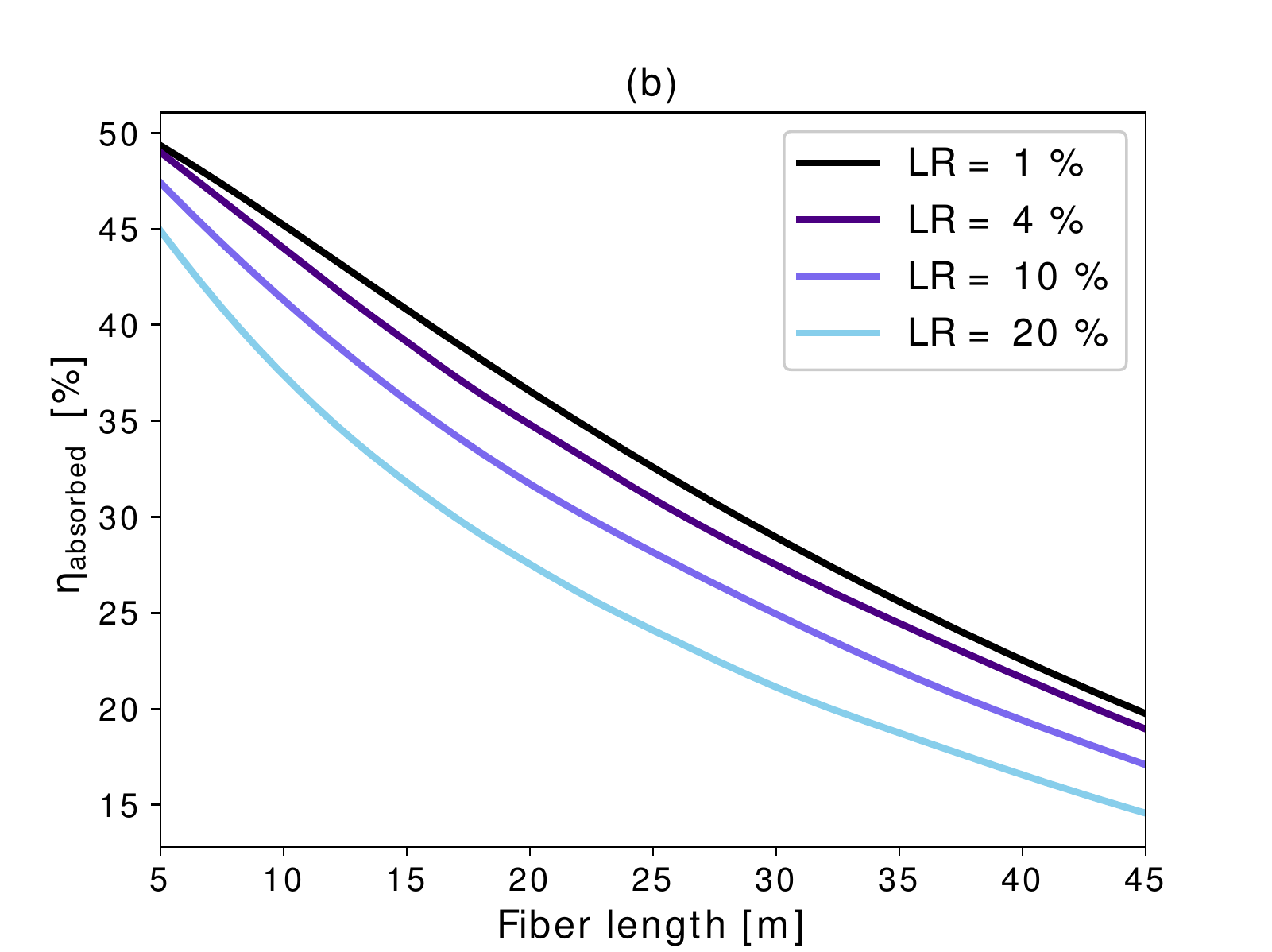}
\end{subfigure}
\caption{(a) Simulation of the output power signal at 1584 nm as a function of the fiber length and the output coupler's reflectivity for an input power of 120 W. (b) Simulation of the efficiency with respect to the absorbed pump power ($\eta_{absorbed}$) as a function of the fiber length and the output coupler's reflectivity.}
\label{fig:simes}
\end{figure}

The characterisation of the chemical and optical properties of the fiber drawn enabled us to perform numerical simulations following the ones described in ref. \cite{Pleau:18} in order to predict the optimal length of the active fiber and the optimal reflectivity of the output coupler. The signal output power at 1584 nm as a function of the fiber length and the output coupler's reflectivity is presented in Fig.\ref{fig:simes}(a). The graph is obtained for an injected pump power of 120 W at 976 nm, which corresponds to the maximal output power of the diode available for the experiments. The value used for the input coupler's reflectivity is 99.9\%. Fig.\ref{fig:simes}(b) presents* the slope efficiency with respect to the absorbed pump power for the same conditions as described for Fig.\ref{fig:simes}(a). 

The simulation results presented in Fig.\ref{fig:simes} give good insight on what to do for an optimal cavity design. Both graphs sketched in Fig.\ref{fig:simes} suggest that the best suited LR-FBG has the lowest reflectivity possible. This means that between 5 and 45 m, the cavity is long enough for the single-pass gain to yield the highest output power possible. The results presented in Fig. \ref{fig:simes}(a) also suggest that in order to reach maximal output power, the length of active fiber should stand between 25 and 30 m. 

\section{Experimental setup}
\begin{figure}[H]
    \centering
    \includegraphics[width=\linewidth]{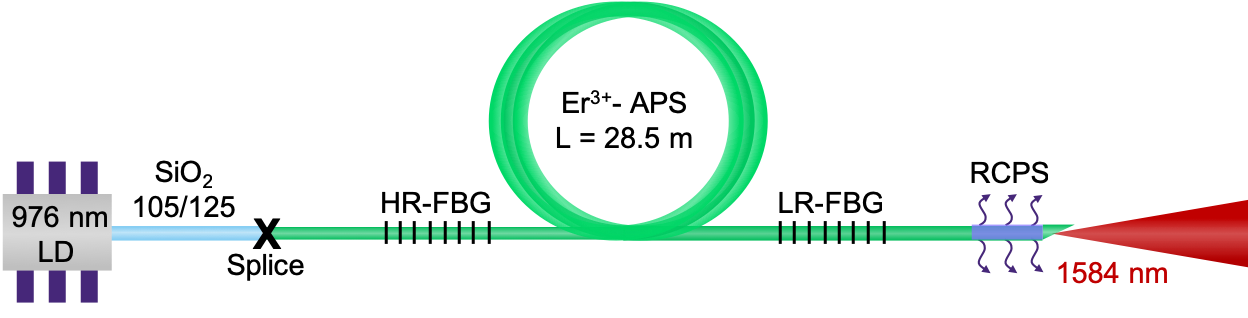}
    \caption{Schematic of the Er$^{3+}$- APS all-fiber laser cavity operating at 1584 nm. The HR (high-reflectivity) and LR (low-reflectivity) FBGs reflect respectively 99.9\% and 1\% of the laser signal. LD: Laser diode. RCPS: residual cladding pump stripper.}
    \label{fig:montage}
\end{figure}

The schematic of the Er$^{3+}$- APS all-fiber laser is depicted in Fig. \ref{fig:montage}. A 976 nm laser diode (nLIGHT element e18) provides the pump signal through a 105/125 µm silica fiber (NA=0.22). The pigtail of the pump diode is joined to the Er$^{3+}$- APS fiber by means of a multi-mode splice performed by a Fujikura arc splicer (model FSM-100P). The active fiber of 28.5 m is bounded by a pair of intracore fiber Bragg gratings, written by femtosecond inscription at 800 nm through the polymer coating \cite{Bernier:14}. Some high index polymer is applied at the end of the cavity along with a 10 cm length of uncoated fiber as a way to strip any residual pump power propagating through the cladding. The output fiber's tip is also cleaved at a 6\si{\degree} angle to prevent parasitic lasing.

\begin{figure}[H]
    \centering
    \includegraphics[width=0.65\linewidth]{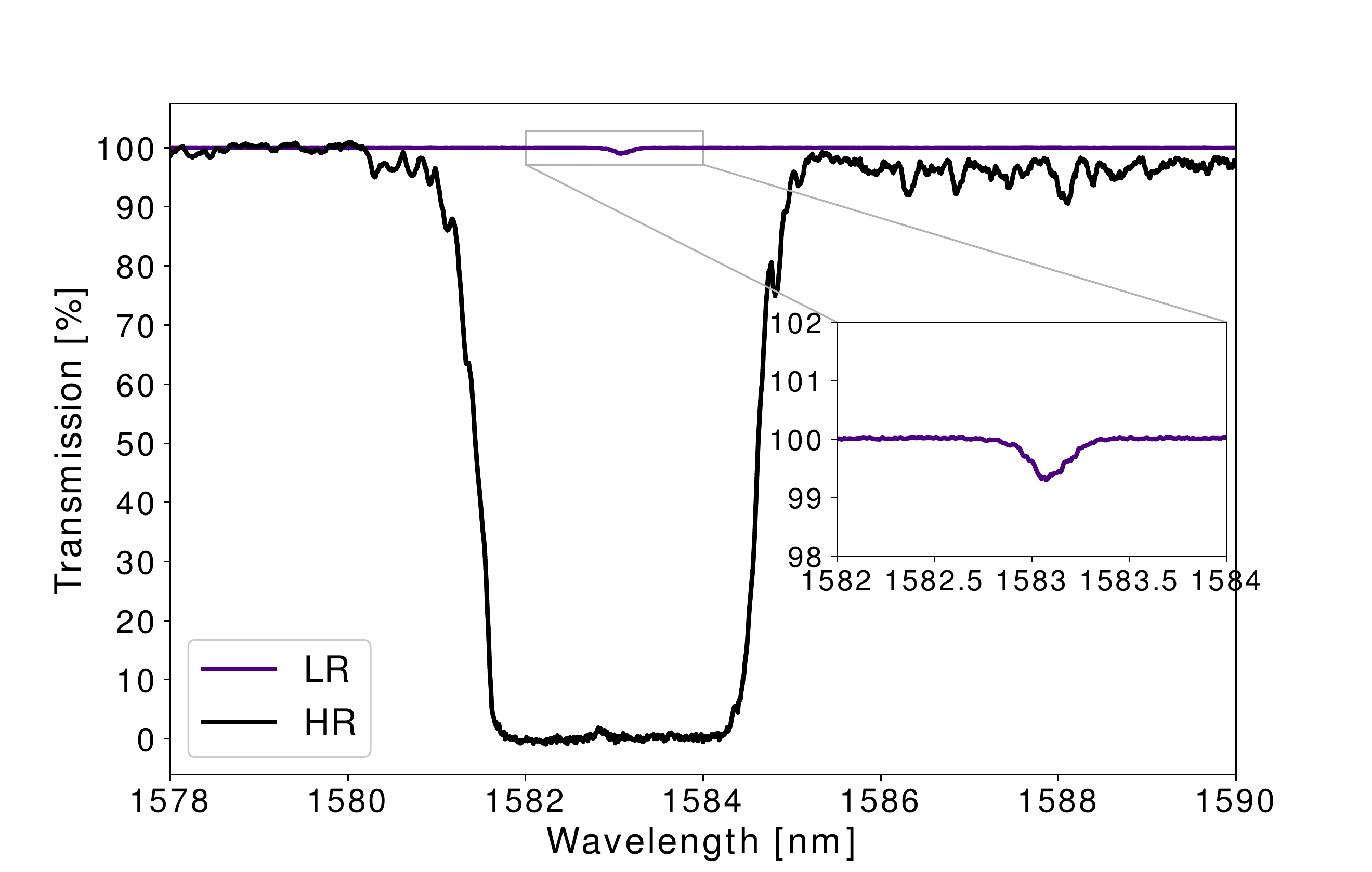}
    \caption{Transmission spectra of the FBGs written in the fiber laser cavity.}
    \label{fig:FBG}
\end{figure}

The spectral transmission of the FBGs is presented in Fig. \ref{fig:FBG}. At 1584 nm, the input coupler (HR-FBG) reflects 99.9\% of signal, while the output coupler (LR-FBG) reflects 1\% of it. 

It should be noted tat the Er$^{3+}$-APS fiber was found to be significantly more photosensitive than the similar designed fiber solely codoped with Al$_2$O$_3$ \cite{Pleau:18}. This allows for the inscription of high performance FBGs, as shown in Fig. \ref{fig:FBG}. Such photosensitivity level could lead to the fabrication of more complex and active FBG structures, such as DFB fiber lasers \cite{loh1998high}.
\section{Results and discussion}
\subsection{Laser performance}

\begin{figure}[H]
\begin{subfigure}{.5\textwidth}
	\centering
\includegraphics[scale=0.45]{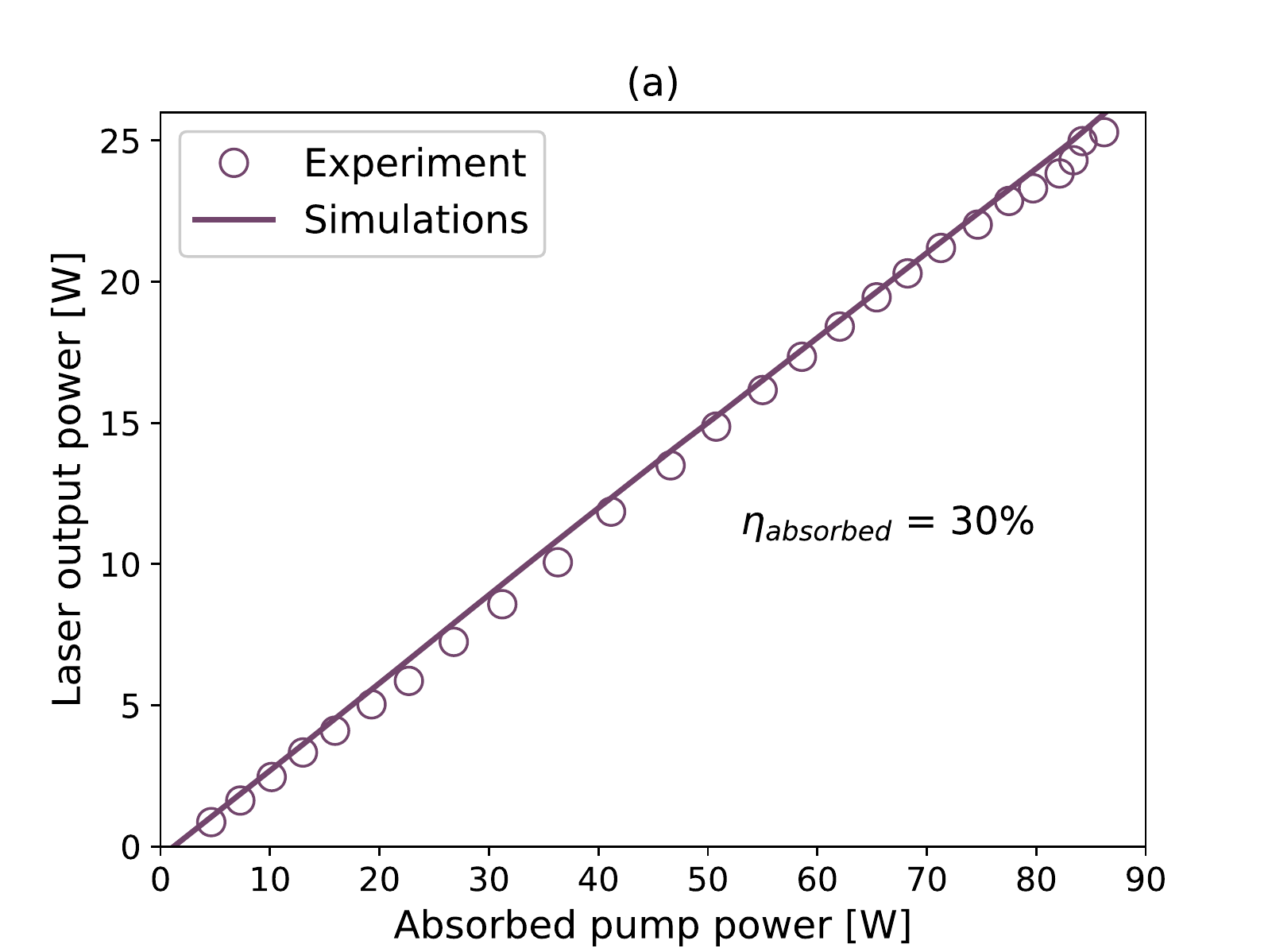}
\end{subfigure}
\begin{subfigure}{.5\textwidth}
	\centering
\includegraphics[scale=0.45]{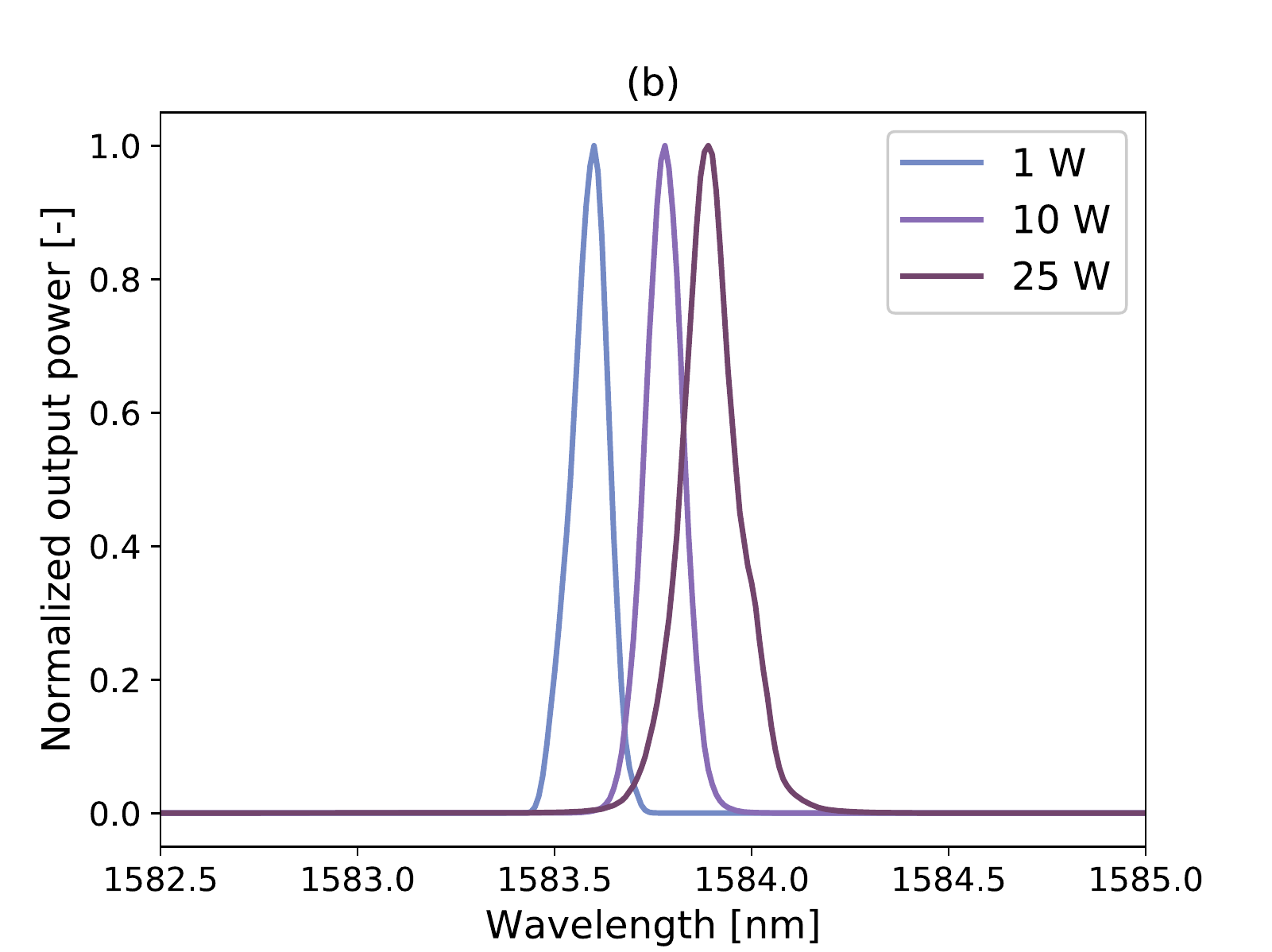}
\end{subfigure}
\caption{(a) Laser output power obtained as a function of absorbed pump power along with the numerical modeling. (b) Spectrum of the laser signal at three different powers.}
\label{fig:laser}
\end{figure}

The output power of the laser signal generated at 1584 nm is presented as a function of the absorbed pump power at 976 nm in Fig. \ref{fig:laser}(a). The experimental results are sketched alongside the theoretical ones obtained from the numerical simulations described previously in section \ref{s:3}. The spectrum of the laser signal at three different output powers is also shown in Fig. \ref{fig:laser}(b).

The slope efficiency with respect to the absorbed pump power at 976 nm is 30\% and the overall efficiency accounting for the launched pump power is 21\%. The maximum output power obtained is 25.3 W for 86.3 W and 121 W of absorbed and launched pump power, respectively. It can be observed from Fig. \ref{fig:laser}(b) that the central wavelength of the laser signal shifts of about 0.3 nm from the lowest to the highest power of operation, demonstrating the stability of the FBGs.

\begin{figure}[H]
\begin{subfigure}{.5\textwidth}
	\centering
\includegraphics[scale=0.44]{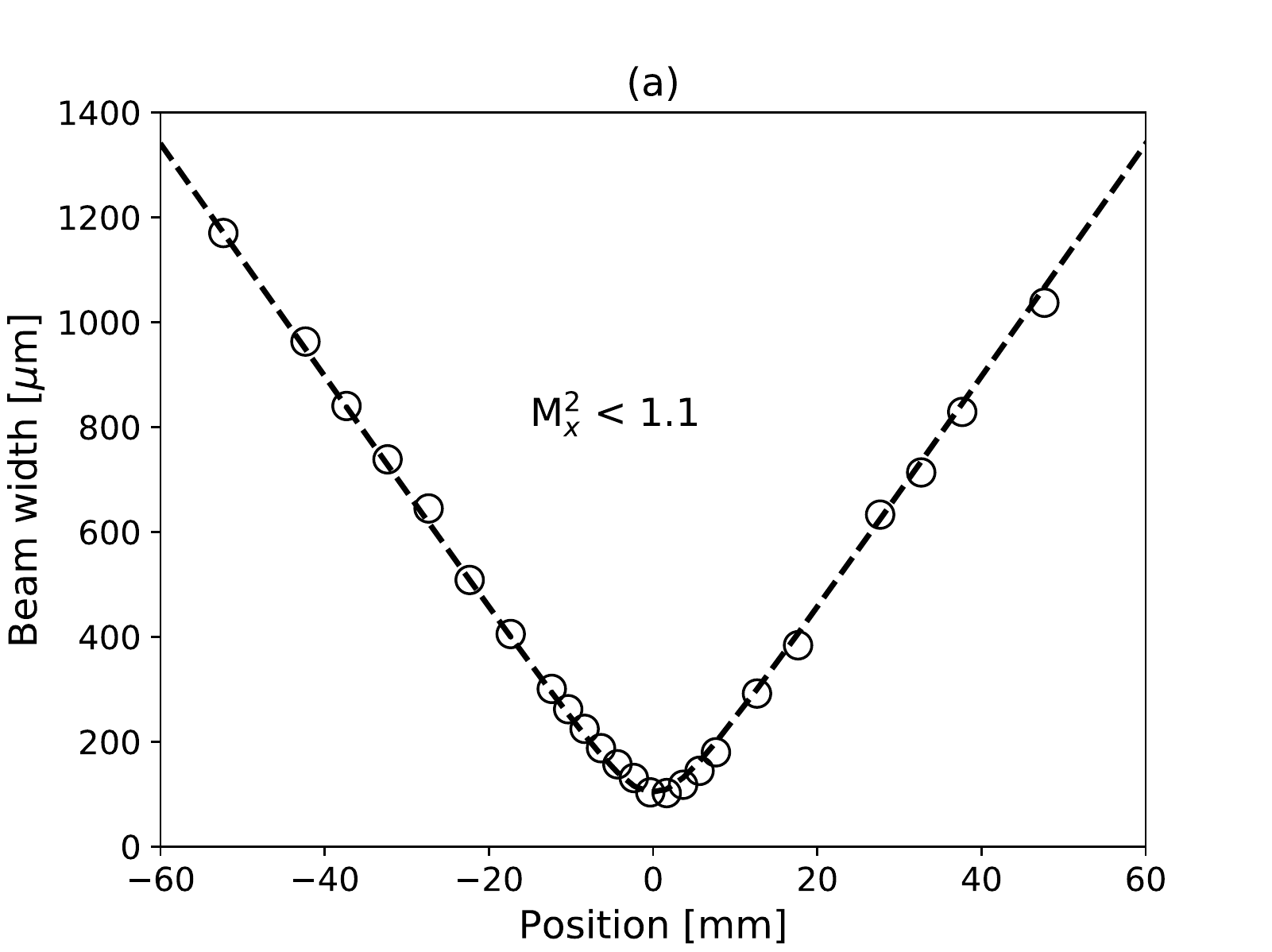}
\label{fig:Mx}
\end{subfigure}
\begin{subfigure}{.5\textwidth}
	\centering
\includegraphics[scale=0.44]{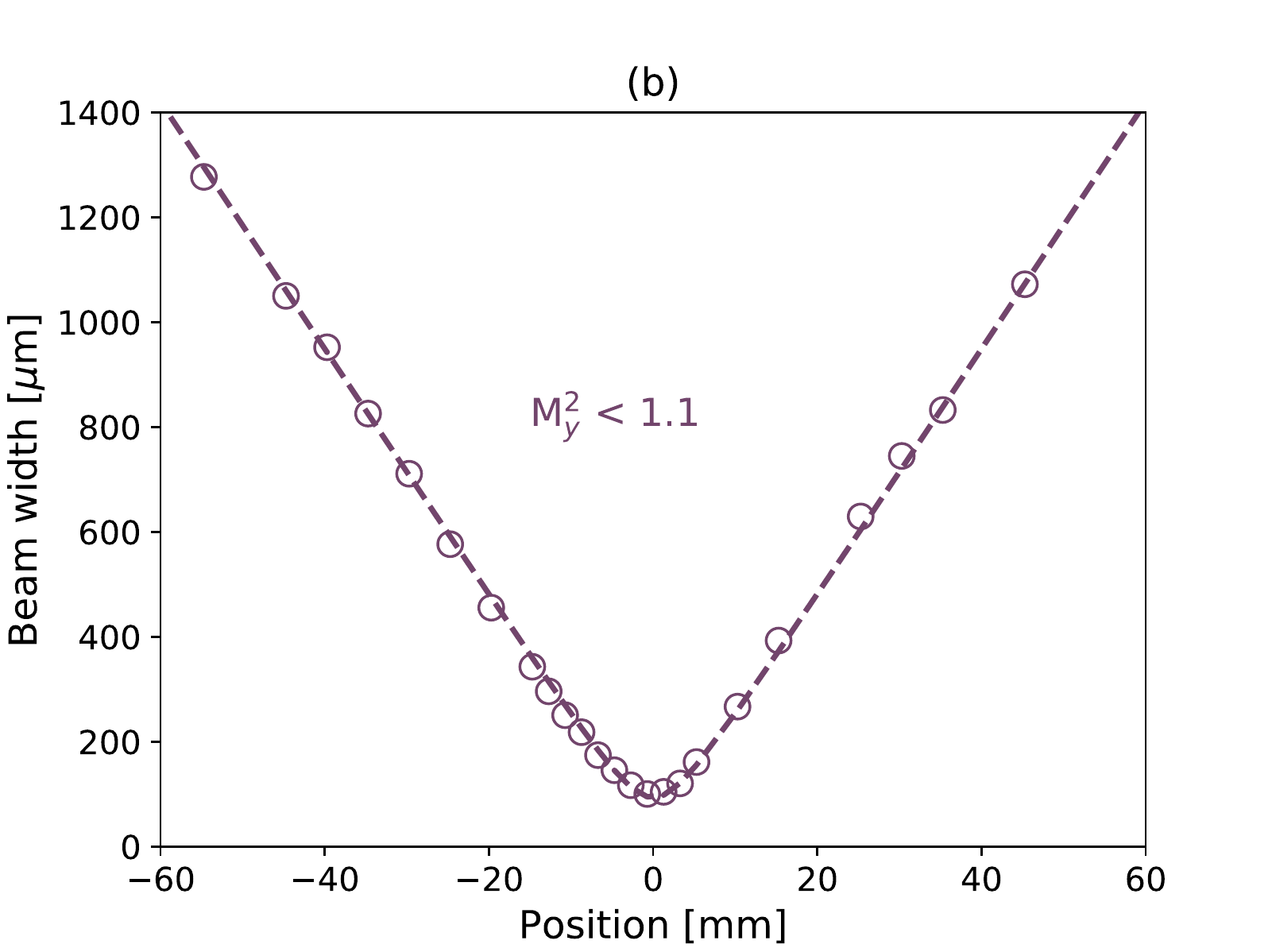}
\label{fig:My}
\end{subfigure}
\caption{Measurement of the beam quality for both axis.}
\label{fig:m2}
\end{figure}

The beam quality of the output signal was measured using an automated bench test (Ophir, Nanoscan). The results are presented in Fig. \ref{fig:m2} and confirm the single-mode operation regime of the laser at 1584 nm (M$^2$ < 1.1). 

\begin{figure}[H]
    \centering
    \includegraphics[scale=0.55]{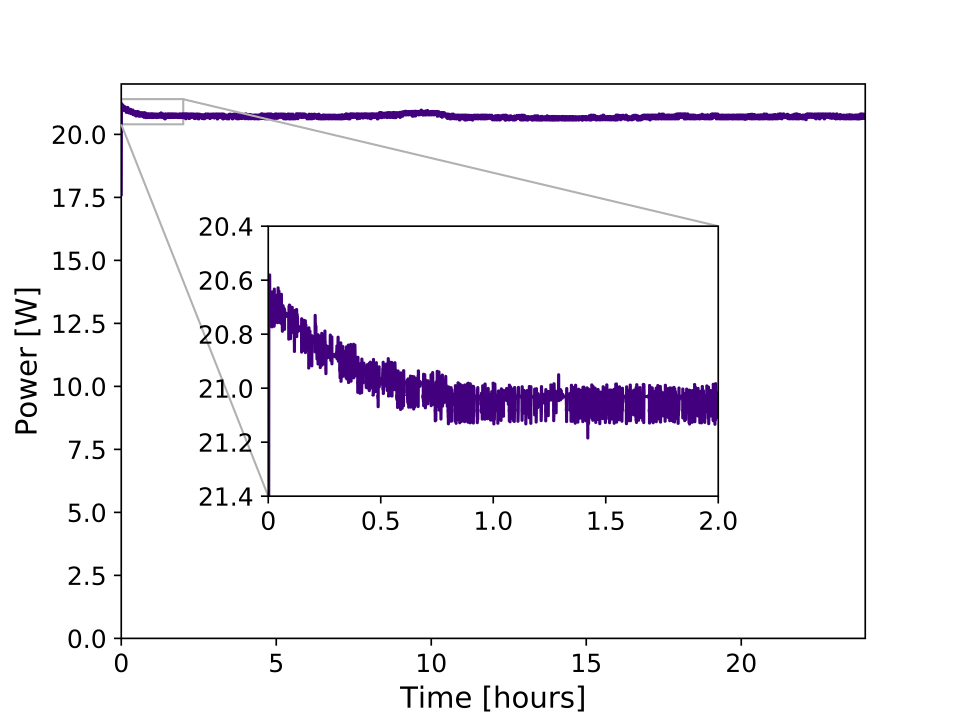}
    \caption{Stability curve of the laser operating at 20 W. The close-up figure shows the power obtained during the warm-up.}
    \label{fig:sta}
\end{figure}

Fig. \ref{fig:sta} presents the laser output power obtained when a constant pump power of 100 W was injected into the cavity for several hours. The results confirm the laser's high stability. 

\subsection{Experimental validation of the numerical model}
On top of the fiber laser described in this paper, four experimental cavities were built in order to test the numerical model's accuracy. They were all constructed with different fiber lengths and were all bounded by a 99.9\% input coupler (HR-FBG) and a 4\% output coupler provided by Fresnel reflection from a straight end cleave. The signal output power was measured for each cavity for an injected pump power of 120 W. 


\begin{figure}[H]
    \centering
    \includegraphics[width=0.6\linewidth]{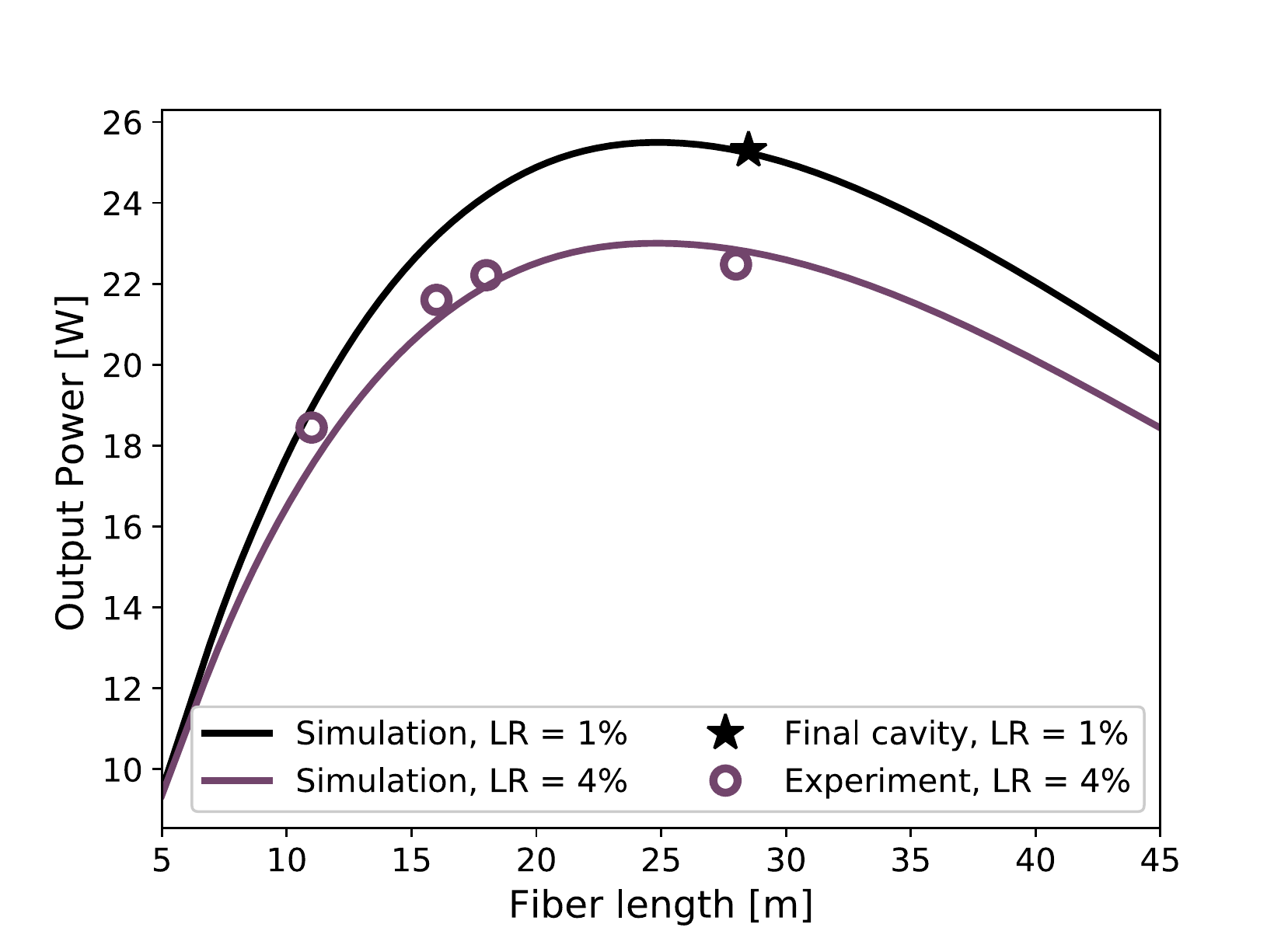}
    \caption{Output power at 1584 nm as a function of the fiber length for an output coupler of 1 and 4 \%. Solid lines represent simulation results, while dots account for experimental data.}
    \label{fig:model}
\end{figure}

\section{Conclusion}

In conclusion, an ytterbium-free erbium-doped aluminophosphosilicate fiber laser is reported. This single-mode all-fiber laser produces over 25 W of stable output power with an efficiency of 30\% with respect to the absorbed pump power provided by a commercial 976 nm diode. The incorporation of aluminium and phosphorus in the fiber core allowed for an increased concentration of erbium (0.06 mol.\%) while keeping both the ion clustering and the numerical aperture to a low value estimated to 4\% and 0.06, respectively. The results show that an optimized fiber recipe based on AlPO\textsubscript{4} would allow for the power scaling of 1.6 µm fiber lasers to many hundred watts in single-mode operation.

\section*{Funding}
Fonds québécois de recherche sur la nature et les technologies (FQRNT); Natural Sciences and Engineering Research Council of Canada (NSERC); Canadian Foundation for Innovation (CFI).

\section*{Acknowledgments}
This research was performed at Université Laval as part of a graduate course (PHY-7052, laboratoire de photonique avancée) taking place during one semester at the department of Physics, Engineering Physics and Optics. The group would like to thank Yannick Ledemi, Stephan Gagnon and Patrick Larochelle for technical assistance.

\section*{Disclosures}
The authors declare no conflicts of interest.

\bibliography{sample}

\begin{thebibliography}{10}
\newcommand{\enquote}[1]{``#1''}

\bibitem{ICNIRP:13}
{International Commission on Non-Ionizing Radiation Protection},
  \enquote{\textsc{ICNIRP} guidelines on limits of exposure to laser radiation
  of wavelengths between 180 nm and 1,000 $\mu$m,}
  {\protect\JournalTitle{Health Phys.}} \textbf{105}, 271--295 (2013).

\bibitem{Frehlich:97}
R.~Frehlich, S.~M. Hannon, and S.~W. Henderson, \enquote{Coherent doppler lidar
  measurements of winds in the weak signal regime,}
  {\protect\JournalTitle{Appl. Opt.}} \textbf{36}, 3491--3499 (1997).

\bibitem{Elgala:11}
H.~Elgala, R.~Mesleh, and H.~Haas, \enquote{Indoor optical wireless
  communication: potential and state-of-the-art,} {\protect\JournalTitle{IEEE
  Communications Magazine}} \textbf{49}, 56--62 (2011).

\bibitem{Solodyankin:08}
M.~A. Solodyankin, E.~D. Obraztsova, A.~S. Lobach, A.~I. Chernov, A.~V.
  Tausenev, V.~I. Konov, and E.~M. Dianov, \enquote{Mode-locked 1.93 $\mu$m
  thulium fiber laser with a carbon nanotube absorber,}
  {\protect\JournalTitle{Opt. Lett.}} \textbf{33}, 1336--1338 (2008).

\bibitem{Delevaque:93}
E.~Delevaque, T.~Georges, M.~Monerie, P.~Lamouler, and J.-F. Bayon,
  \enquote{Modeling of pair-induced quenching in erbium-doped silicate fibers,}
  {\protect\JournalTitle{IEEE Photonics Technology Letters}} \textbf{5}, 73--75
  (1993).

\bibitem{Lin:18}
H.~Lin, Y.~Feng, Y.~Feng, P.~Barua, J.~K. Sahu, and J.~Nilsson, \enquote{656 w
  er-doped, yb-free large-core fiber laser,} {\protect\JournalTitle{Opt.
  Lett.}} \textbf{43}, 3080--3083 (2018).

\bibitem{Jebali:14}
M.~A. Jebali, J.-N. Maran, and S.~LaRochelle, \enquote{264 w output power at
  1585 nm in er–yb codoped fiber laser using in-band pumping,}
  {\protect\JournalTitle{Opt. Lett.}} \textbf{39}, 3974--3977 (2014).

\bibitem{Jeong:07}
Y.~Jeong, S.~Yoo, C.~A. Codemard, J.~Nilsson, J.~K. Sahu, D.~N. Payne,
  R.~Horley, P.~W. Turner, L.~Hickey, A.~Harker, M.~Lovelady, and A.~Piper,
  \enquote{Erbium:ytterbium codoped large-core fiber laser with 297-w
  continuous-wave output power,} {\protect\JournalTitle{IEEE Journal of
  Selected Topics in Quantum Electronics}} \textbf{13}, 573--579 (2007).

\bibitem{Scrivener:89}
P.~Scrivener, E.~Tarbox, and P.~Maton, \enquote{Narrow linewidth tunable
  operation of er/sup 3+/-doped single-mode fibre laser,}
  {\protect\JournalTitle{Electronics Letters}} \textbf{25}, 549--550 (1989).

\bibitem{Vienne:98}
G.~G. Vienne, J.~E. Caplen, L.~Dong, J.~D. Minelly, J.~Nilsson, and D.~N.
  Payne, \enquote{Fabrication and characterization of yb3+: Er3+
  phosphosilicate fibers for lasers,} {\protect\JournalTitle{Journal of
  lightwave technology}} \textbf{16}, 1990 (1998).

\bibitem{Funabiki:12}
F.~Funabiki, T.~Kamiya, and H.~Hosono, \enquote{Doping effects in amorphous
  oxides,} {\protect\JournalTitle{Journal of the Ceramic Society of Japan}}
  \textbf{120}, 447--457 (2012).

\bibitem{Savelii2017}
I.~Savelii, L.~Bigot, B.~Capoen, C.~Gonnet, C.~Chan{\'e}ac, E.~Burova,
  A.~Pastouret, H.~El-Hamzaoui, and M.~Bouazaoui, \enquote{Benefit of
  rare-earth ``smart doping'' and material nanostructuring for the next
  generation of er-doped fibers,} {\protect\JournalTitle{Nanoscale Research
  Letters}} \textbf{12}, 206 (2017).

\bibitem{Likhachev:09}
M.~E. Likhachev, M.~M. Bubnov, K.~V. Zotov, D.~S. Lipatov, M.~V. Yashkov, and
  A.~N. Guryanov, \enquote{Effect of the alpo 4 join on the pump-to-signal
  conversion efficiency in heavily er-doped fibers,}
  {\protect\JournalTitle{Optics letters}} \textbf{34}, 3355--3357 (2009).

\bibitem{aboud2015}
T.~Aboud, \enquote{Aluminophosphosilicate glasses as an alternative source for
  microporous materials of technically attractive compositions and
  morphological characteristics,} in \emph{Mechanical and Electrical Technology
  VII,}  vol. 799 of \emph{Applied Mechanics and Materials} (Trans Tech
  Publications Ltd, 2015), pp. 145--152.

\bibitem{Liu:18}
S.~Liu, K.~Peng, H.~Zhan, L.~Ni, X.~Wang, Y.~Wang, Y.~Li, J.~Yu, L.~Jiang,
  R.~Zhu, J.~Wang, F.~Jing, and A.~Lin, \enquote{3 kw 20/400 yb-doped
  aluminophosphosilicate fiber with high stability,}
  {\protect\JournalTitle{IEEE Photonics Journal}} \textbf{10}, 1--8 (2018).

\bibitem{kotov2012high}
L.~V. Kotov, M.~E. Likhachev, M.~M. Bubnov, O.~I. Medvedkov, D.~S. Lipatov,
  N.~N. Vechkanov, and A.~N. Guryanov, \enquote{High-performace cladding-pumped
  erbium-doped fibre laser and amplifier,} {\protect\JournalTitle{Quantum
  Electronics}} \textbf{42}, 432 (2012).

\bibitem{Pleau:18}
L.-P. Pleau, P.~Paradis, J.-S. Freni{\'e}re, M.~Huneault, S.~Gouin, S.~M.
  Aljamimi, Y.~O. Aydin, S.~Duval, J.-C. Gauthier, J.~Habel \emph{et~al.},
  \enquote{20 w splice-free erbium-doped all-fiber laser operating at 1610 nm,}
  {\protect\JournalTitle{Optics express}} \textbf{26}, 22378--22388 (2018).

\bibitem{Barnes:91}
W.~L. Barnes, R.~I. Laming, E.~J. Tarbox, and P.~R. Morkel, \enquote{Absorption
  and emission cross section of er$^{3+}$ doped silica fibers,}
  {\protect\JournalTitle{IEEE Journal of Quantum Electronics}} \textbf{27},
  1004--1010 (1991).

\bibitem{Bernier:14}
M.~Bernier, F.~Tr\'{e}panier, J.~Carrier, and R.~Vall\'{e}e, \enquote{High
  mechanical strength fiber bragg gratings made with infrared femtosecond
  pulses and a phase mask,} {\protect\JournalTitle{Opt. Lett.}} \textbf{39},
  3646--3649 (2014).

\bibitem{loh1998high}
W.~Loh, B.~Samson, L.~Dong, G.~Cowle, and K.~Hsu, \enquote{High performance
  single frequency fiber grating-based erbium: ytterbium-codoped fiber lasers,}
  {\protect\JournalTitle{Journal of lightwave technology}} \textbf{16}, 114
  (1998).

\end{thebibliography}

\end{document}